\documentclass{article}
\usepackage{spconf,amsmath,graphicx}
\usepackage{amsfonts}
\usepackage{xcolor}
\usepackage{color}
\usepackage{graphicx}
\usepackage{subcaption}
\usepackage{multirow}
\usepackage{cite}
\usepackage{multicol}
\usepackage{booktabs}

\title{Sparsely Shared LoRA on Whisper for Child Speech Recognition}
\name{Wei Liu$^{1}$, Ying Qin$^{2}$, Zhiyuan Peng$^{1}$, Tan Lee$^{1}$ }
\address{
  $^1$ Department of Electronic Engineering, The Chinese University of Hong Kong  \\
  $^2$ Institute of Information Science, Beijing Jiaotong University, Beijing 100044, China\\ 
  }
\begin{document}
\ninept
\maketitle
\begin{abstract}


Whisper is a powerful automatic speech recognition (ASR) model. Nevertheless, its zero-shot performance on low-resource speech requires further improvement. Child speech, as a representative type of low-resource speech, is leveraged for adaptation.
Recently, parameter-efficient fine-tuning (PEFT) in NLP was shown to be comparable and even better than full fine-tuning, while only needing to tune a small set of trainable parameters. However, current PEFT methods have not been well examined for their effectiveness on Whisper. In this paper, only parameter composition types of PEFT approaches such as LoRA and Bitfit are investigated as they do not bring extra inference costs.
Different popular PEFT methods are examined. Particularly, we compare LoRA and AdaLoRA and figure out the learnable rank coefficient is a good design. Inspired by the sparse rank distribution allocated by AdaLoRA, a novel PEFT approach Sparsely Shared LoRA (S2-LoRA) is proposed. The two low-rank decomposed matrices are globally shared. Each weight matrix only has to maintain its specific rank coefficients that are constrained to be sparse. 
Experiments on low-resource Chinese child speech show that with much fewer trainable parameters, S2-LoRA can achieve comparable in-domain adaptation performance to AdaLoRA and exhibit better generalization ability on out-of-domain data. In addition, the rank distribution automatically learned by S2-LoRA is found to have similar patterns to AdaLoRA's allocation. 

\end{abstract}
\section{Introduction}
\label{sec:intro}

Automatic speech recognition (ASR) nowadays has become very powerful. Benefiting from large amounts of training data \cite{baevski2020wav2vec, DBLP:conf/interspeech/BabuWTLXGSPSPBC22} 
, representative ASR models, like Whisper \cite{radford2023robust}, USM \cite{zhang2023google}, and MMS \cite{pratap2023scaling}, demonstrate remarkable multi-task and multi-lingual recognition capabilities. It is claimed that Whisper can perform zero-shot speech recognition across different languages and domains. 
Meanwhile, recognition of low-resource speech, such as child speech, disorder speech, and speech from the endangered languages, has persistently presented formidable challenges \cite{shivakumar2022end, ye2021development, yi2020applying}. Consequently, it becomes imperative to examine the zero-shot performance of Whisper on these data. More importantly, should a performance gap be identified, it is essential to explore efficient adaptation strategies.
Chinese child speech, as a representative type of low-resource speech, is leveraged in this work to probe the performance of Whisper and explore effective adaptation approaches. 



The common practice for adaptation is to collect in-domain data and conduct model fine-tuning. 
However, two widely concerned issues exist. (1) The target domain data for low-resource speech is very limited. (2) With the larger model, full fine-tuning becomes increasingly computational costing, and the fine-tuned model tends to overfit. 
Recently, parameter-efficient fine tuning (PEFT) has attained great attention in the NLP community \cite{ruder2022modular, ding2023parameter, DBLP:conf/iclr/HeZMBN22}, mainly for large language models (LLMs). PEFT approaches can be roughly divided into three categories \cite{ruder2022modular}. (a) \textit{function composition}: like adapter \cite{houlsby2019parameter}; (b) \textit{input composition}: like prompt tuning and prefix tuning \cite{DBLP:conf/emnlp/LesterAC21, DBLP:conf/acl/LiL20}; (c) \textit{parameter composition}: Low-rank adaptation (LoRA) \cite{DBLP:conf/iclr/HuSWALWWC22}, BitFit \cite{DBLP:conf/acl/ZakenGR22}, etc \cite{DBLP:conf/iclr/ZhangCBH0CZ23, chavan2023one, liu2022few}. These approaches introduce a small proportion of learnable parameters in different forms to the original model. Adapters directly insert lightweight neural network modules into the pre-trained model, which changes the model architecture. Prompt tuning and prefix tuning add leading learnable embeddings to the input sequence, resulting in the increased inference cost due to longer input sequences. Different from adapters and prompt tuning, 
\textit{parameter composition}'s PEFT carries no additional inference cost once it has finished training. This type of approach enables the trainable parameters to be seamlessly integrated into the original weights, thereby promoting \textit{parameter composition} as a prevailing fine-tuning strategy.

\begin{figure*}[th!]
  \centering
  \vspace{-5mm}
  \includegraphics[width=0.8\linewidth]{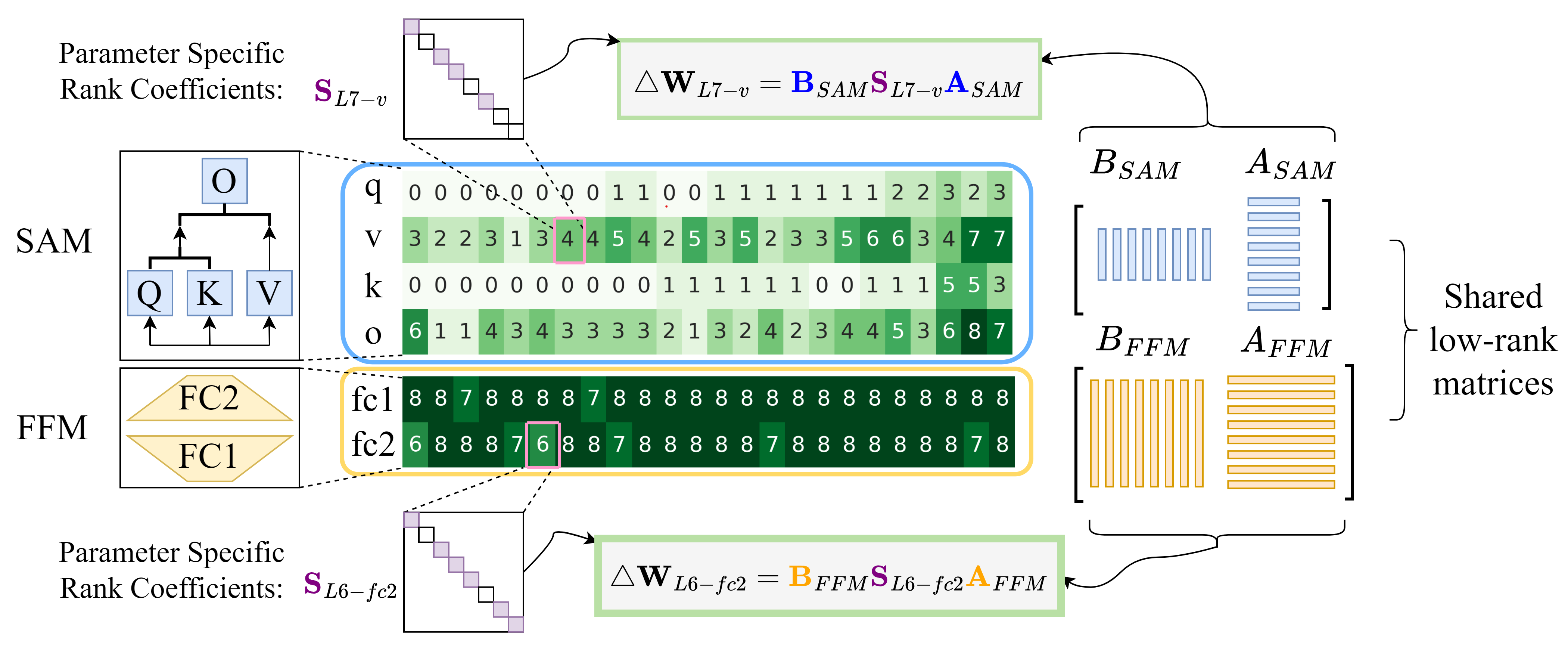}
  \caption{The overall illustration of S2-LoRA. Numbers in the green matrices denote the ranks distributed across all weight matrices of the Whisper encoder. The maximum shared rank is 8. The Whisper \textit{medium}'s encoder with 24 layers is shown as an example where SAM and FFM denote self-attention and feed-forward modules, the decoder operates similarly while including another cross-attention module (CAM). }
  \label{fig:S2-LoRA}
  \vspace{-3mm}
\end{figure*}

LoRA, as the most representative \textit{parameter composition} approach, has several variants, particularly, AdaLoRA \cite{DBLP:conf/iclr/ZhangCBH0CZ23}. 
It is known that LoRA manually specifies a fixed rank for all the weight matrices, overlooking the varying importance between different weights. AdaLoRA addresses the issue with an importance-aware rank allocator. It maintains an overall rank budget and dynamically distributes the ranks to weight matrices according to an importance metric. 

In this paper, several research questions are raised and addressed: (1) Why do we need PEFT? The effectiveness of these \textit{parameter composition} types of PEFT methods are examined on Whisper adaptation to child speech. (2) What makes AdaLoRA superior to LoRA? A series of ablation studies are conducted to verify the importance of the rank coefficient design. (3) To what level the trainable parameters can be reduced while keeping performance?  Inspired by the sparsity property of AdaLoRA's rank distribution, the Sparsely Shared LoRA (S2-LoRA) is proposed. The two low-rank matrices of LoRA are module-wise globally shared. Each weight matrix within the module only needs to adjust its specific sparse coefficient vector to control each rank's importance, achieving the same purpose of adaptive rank allocation but with much fewer parameters (only $0.02\%$). Experiments on low-resource Chinese child speech further demonstrate the effectiveness of S2-LoRA.


\section{S2-LoRA on Whisper model}
\label{sec:methods}
This section starts with a brief introduction of Whisper, followed by the formulations of LoRA, AdaLoRA, and the proposed S2-LoRA.

\subsection{Whisper}
Whisper \cite{radford2023robust} has a typical transformer-based encoder-decoder architecture that is designed for multiple speech tasks, including multilingual ASR, multilingual speech translation, language identification, etc. 
The transformer \cite{vaswani2017attention} structure consists of three submodules as building blocks, namely the self-attention module (SAM), feed-forward module (FFM), and cross-attention module (CAM). The encoder layer comprises SAM and FFM, while the decoder layer includes all of the three. The learnable weight matrices in both SAM and CAM are $\{\mathbf{W_q}, \mathbf{W}_k, \mathbf{W}_v, \mathbf{W}_o\}$, which represents the query, key, value, and output projection matrices, respectively. In FFM, $\{\mathbf{W}_{fc1}, \mathbf{W}_{fc2}\}$ are the projection matrices of the fully-connected layers.

Whisper is trained on around 680,000 hours of weakly-supervised speech data collected from the Internet, with the performance approaching human-level accuracy and robustness.



\subsection{LoRA}
LoRA \cite{DBLP:conf/iclr/HuSWALWWC22} was first proposed in NLP to efficiently adapt large language models (LLMs) to specific domains or downstream tasks. 
It was found that the weights of pre-trained LLMs tend to reside in a low intrinsic dimensional space \cite{DBLP:conf/acl/AghajanyanGZ20}. Inspired by the observation, LoRA freezes the original weights and only updates the low-rank incremental weight matrices. 
Specifically, consider the $i$-th linear projection forward, $f_{i}(\mathbf{x}) = \mathbf{x}\mathbf{W}_i^{T}+ \mathbf{b}_{i}$, where $\mathbf{W}_i \in \mathbb{R}^{d_2\times d_1}$ and $\mathbf{b}_i \in \mathbb{R}^{d_2}$ denotes the frozen weight and bias. By applying LoRA, the forward is modified as,
\begin{equation}
    f_i(\mathbf{x}) = \mathbf{x}(\mathbf{W}_i + \Delta\mathbf{W}_{i})^{T} + \mathbf{b}_{i}; ~~\Delta\mathbf{W}_i = \mathbf{B}_i\mathbf{A}_i,
\end{equation}
 where $\mathbf{B}_i \in \mathbb{R}^{d_2 \times r}$ and $\mathbf{A}_i \in \mathbb{R}^{r \times d_1}$ are the two trainable rank-decomposed matrices, with the rank $r \ll \min\{d_1, d_2\}$. 

\subsection{AdaLoRA}
\label{subsec:adalora}
The importance of weight parameters to the performance varies across different layers and modules. Intuitively, some weight matrices should be allocated with higher ranks than others in adaptation. However,
LoRA specifies a fixed rank for all the weight matrices, which overlooks the varying importance of weights and can be sub-optimal. 
AdaLoRA \cite{DBLP:conf/iclr/ZhangCBH0CZ23} addresses this issue with an importance-aware rank allocation method. 
It comes with two modifications to LoRA. First, the incremental update $\Delta\mathbf{W}_i$ is parameterized in the form of singular value decomposition (SVD), i.e.,
\begin{equation}
\label{eq:adalora}
    \Delta\mathbf{W}_i = \mathbf{\bar{B}}_i\mathbf{\Lambda}_i\mathbf{\bar{A}}_i; ~~\mathbf{\bar{B}}_i^{T}\mathbf{\bar{B}}_i=\mathbf{\bar{A}}_i\mathbf{\bar{A}}_i^{T} = \mathbf{I},
\end{equation}
where $\mathbf{\bar{B}}_i \in \mathbb{R}^{d_2 \times r}$ and $\mathbf{\bar{A}}_i \in \mathbb{R}^{r \times d_1}$ contain the left/right singular vectors of $\Delta\mathbf{W}_i$ . $\mathbf{I}$ denotes the identity matrix. and the $\{\mathbf{\bar{B}}_i, \mathbf{\bar{A}}_i\}$ are enforced to be orthogonal. The diagonal matrix $\mathbf{\Lambda}_i \in \mathbb{R}^{r\times r}$ contains $r$ singular values, with $r \ll \min\{d_1, d_2\}$.  Only $\mathbf{\bar{B}}_i, \mathbf{\bar{A}}_i, \mathbf{\Lambda}_i$ are trainable.

Second, in adaptation, AdaLoRA dynamically allocates an overall rank budget to its update matrices $\{\Delta \mathbf{W}_i\}$. This is achieved by iteratively masking out less important singular values after every gradient update step. A sensitivity-based importance metric \cite{zhang2022platon} is utilized to measure and sort the importance of the $k$-th triplet $\{\mathbf{\Lambda}_{i}^{k,k}, \mathbf{\bar{B}}^{*,k}_{i}, \mathbf{\bar{A}}^{k,*}_{i}\}$ of the $i$-th weight matrix $\mathbf{W}_i$, which takes account of both singular values and vectors. The non-zero $\mathbf{\Lambda}_{i}^{k,k}$ acts as the rank coefficient to control the allocated rank budget.


\subsection{Proposed S2-LoRA}



Our preliminary experiments on AdaLoRA (See Fig. \ref{fig:s2-lora-rank}) show that (1) it tends to allocate significantly fewer rank budgets to $\{\mathbf{W}_q$,$\mathbf{W}_k\}$  than those to $\{\mathbf{W}_v$,$\mathbf{W}_o\}$ in both SAM and CAM; (2) the sparsity of rank budgets among layers exists. In this regard, parameters for adaptation can be further reduced. It may not be necessary to learn the rank-decomposed matrices $\mathbf{B}_i, \mathbf{A}_i$ separately for each weight matrix $\mathbf{W}_i$. They can be tied and shared across layers and modules.
A novel PEFT approach, \textbf{Sparsely Shared Low-rank Adaptation} (S2-LoRA), is therefore proposed in this work. 
S2-LoRA has $\mathbf{B}$ and $\mathbf{A}$ globally shared. The $i$-th weight matrix only stores a single trainable rank coefficient vector $\mathbf{s}_i$. In addition, S2-LoRA introduces an L1 constraint on the rank vector for sparsity. The orthogonal constraint proposed in Eq. \ref{eq:adalora} is discarded. To this end, the incremental update for each weight matrix $\mathbf{W}_i$ is given as:
\begin{equation}
    \Delta\mathbf{W}_i = \mathbf{B}\mathbf{S}_i\mathbf{A}; ~~\text{s.t.}~~ ||\mathbf{S}_i||_{1} < \epsilon,
\end{equation}
where $\mathbf{S}_i = diag(\mathbf{s}_i)$ and $\epsilon$ denotes an empirically small value.
The objective function of S2-LoRA can be written as:
\begin{align}
\begin{split}
\label{eq:loss}
    \mathcal{L}_{total} & = \mathcal{L}_{\text{CE}}(\mathcal{D}|\mathbf{B},\mathbf{A}, \{\mathbf{S}_i\}_{i=1}^{N}) + \alpha_1 * \frac{1}{N} \sum_{i=1}^{N}||\mathbf{S}_i||_{1} \\
    & + \alpha_2 * \frac{1}{2r} \sum_{k=1}^{r}(||\mathbf{B}^{*,k}||_2 + ||\mathbf{A}^{k,*}||_2),
\end{split}
\end{align}
where $\mathcal{L}_{\text{CE}}$ represents the original cross-entropy loss for fine-tuning Whisper. $\mathcal{D}$ is the adaptation data. $N$ denotes the number of trainable weight matrices. $\alpha_1$ controls the sparsity penalty on $\mathbf{S}_i$. $\alpha_2$ regularizes $\{\mathbf{B}, \mathbf{A} \}$ by L2 norm, avoiding the numeric scaling issue in training.

In practice, as shown in Fig. \ref{fig:S2-LoRA}, the $\{\mathbf{B}, \mathbf{A} \}$ of S2-LoRA are not globally shared across the whole Whisper model. 
The functionality of the encoder and decoder in Whipser are different. The three basic blocks, SAM, CAM, and FFM, act as various modeling roles. 
Consequently, we split Whisper into 5 modules, namely, {Enc-SAM, Enc-FFM, Dec-SAM, Dec-CAM, and Dec-FFM}. $\{\mathbf{B}, \mathbf{A} \}$ are tied only within each module. Fig. \ref{fig:S2-LoRA} gives the details of how S2-LoRA is applied to Enc-SAM and Enc-FFM. 

\section{Experiments Setup}
\label{sec:exp_setup}
\subsection{Datasets}
The Chinese child speech for Whisper adaptation comes from the subsets of CSRC-2021 dataset \footnote{https://www.data-baker.com/csrc\_challenge.html} \cite{yu2021slt}. CSRC-2021 contains 28.6 hours of child-read speech (zh-C1) and 29.5 hours of child conversational speech (zh-C2). The zh-C1 is the training set in the target domain, which has around 30K utterances. To investigate the effect of different amounts of adaptation data, the zh-C1 was split into three sets: namely zh-C1-1K, zh-C1-10K, and the original zh-C1-30K. In the evaluation stage, three Chinese Mandarin test sets were prepared. They are child read/conversational speech from the evaluation sets of CSRC-2021 and adult read speech from Aishell1 \cite{bu2017aishell} test set (zh-A1). 
In addition, Cantonese (zh-HK) and English (en) from Common Voice 11.0 \cite{DBLP:conf/lrec/ArdilaBDKMHMSTW20} test sets are utilized to evaluate the cross-lingual generalization.  
Each of the above test sets contains the randomly selected 1,000 utterances for fast evaluation. 

\subsection{PEFT Configurations}
The PEFT methods we compared belong to the category of \textit{parameter composition}. 
(1) BitFit \cite{DBLP:conf/acl/ZakenGR22}: update all the bias parameters of the model. (2) IA3 \cite{liu2022few}: introduce three learned vectors per layer to elementwisely multiply with the projection output from $\{\mathbf{W}_{k}, \mathbf{W}_v, \mathbf{W}_{fc2}\}$. (3) LoRA \cite{DBLP:conf/iclr/HuSWALWWC22}: model incremental updates of  $\{\mathbf{W}_{q}, \mathbf{W}_v\}$ with the low-rank matrices where rank $r=8$. (4) GLoRA \cite{chavan2023one}: a generalized LoRA that gives a unified formulation encompassing all tunable dimensions. In GLoRA, $\Delta\mathbf{W}_i = \mathbf{W}_i\mathbf{A}_i + \mathbf{B}_i$, $\Delta\mathbf{b}_i = \mathbf{C}_i\mathbf{W}_i + \mathbf{D}_i\mathbf{b}_i + \mathbf{E}_i$, where $\{\mathbf{A}_i, \mathbf{B}_i, \mathbf{C}_i, \mathbf{D}_i, \mathbf{E}_i\}$ are trainable parameters that can be further low-rank decomposed for matrices ($r=8$). Note that GLoRA becomes LoRA if only having  $\mathbf{B}$ and becomes BitFit if only $\mathbf{E}$ exists. (5) AdaLoRA \cite{DBLP:conf/iclr/ZhangCBH0CZ23}: dynamically allocate the rank budget to weight matrices of $\{\mathbf{W}_{q}, \mathbf{W}_v\}$, where the initial rank is set to 12 and target rank is 8. (6) S2-LoRA: propose to sparsely share the low-rank components ($r=8$), every weight matrix only need to maintain a rank coefficient vector $\mathbf{s}$. In Eq. \ref{eq:loss}, the sparsity weight $\alpha_1$ and regularization weight $\alpha_2$ is set to $0.05$ and $0.1$ respectively. All the above methods are implemented with the open-sourced Huggingface PEFT library \footnote{https://github.com/huggingface/peft}.

\subsection{Whisper Tuning}
Whisper has five different sizes, namely \textit{tiny} (39M), \textit{base} (74M), \textit{small} (244M), \textit{medium} (769M), and \textit{large} (1.5B). The Huggingface trainer is used to fine-tune the model. Since the speech input to Whisper would be padded to 30 seconds, $batch\_size$=2 is used with 8 gradient accumulations. Mixed precision training is enabled. The number of epochs is set to 3. The learning rate is 1e-3 in PEFT mode while using 1e-4 under full fine-tuning. Greedy decoding is applied for ASR inference. Character error rate (CER) is used to evaluate Chinese languages and word error rate (WER) is used for English. 

\begin{figure}[t!]
  \centering
  \includegraphics[width=\linewidth]{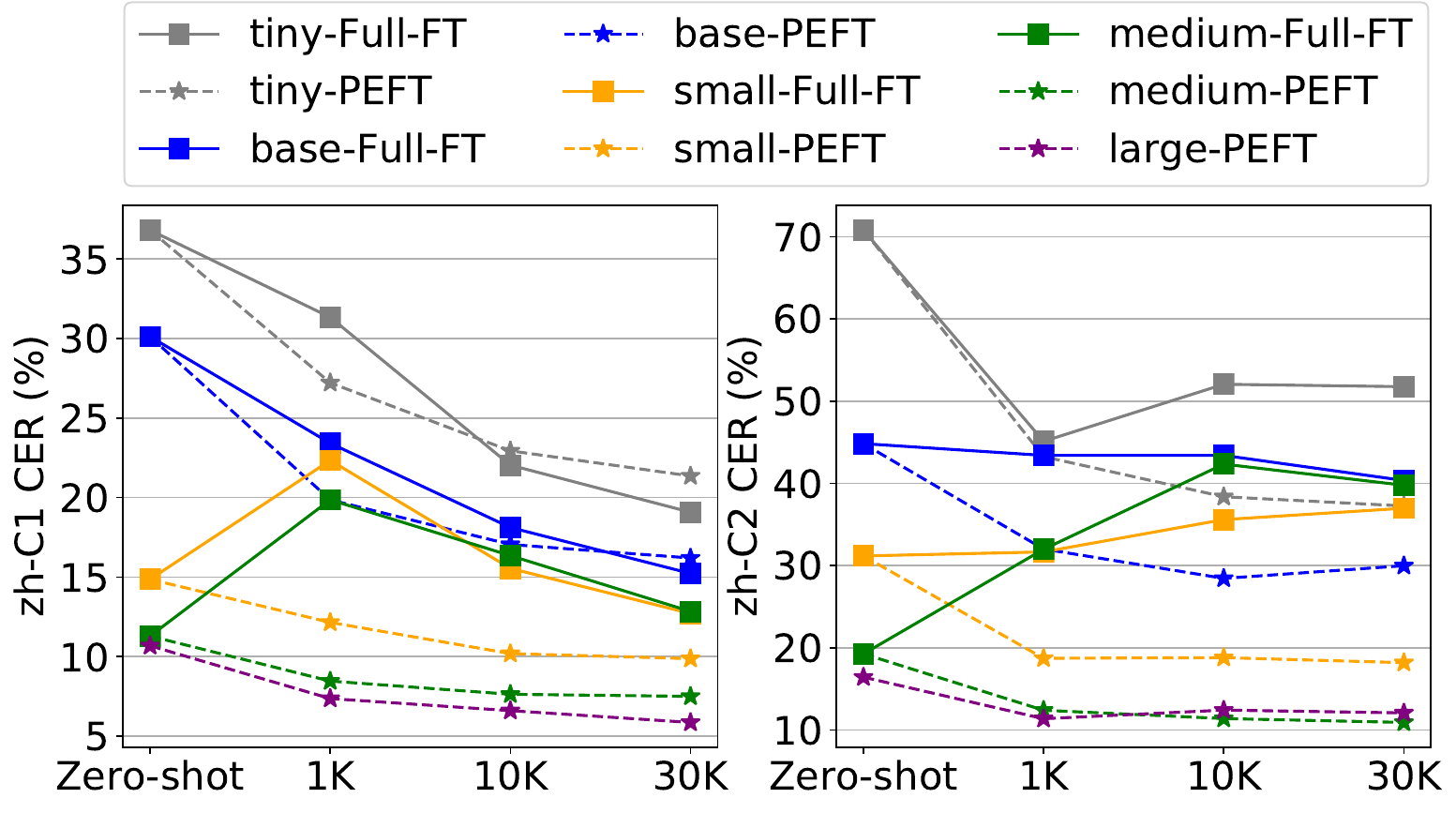}
  \caption{Full-FT vs. PEFT on Whisper with different amount of adaptation data zh-C1. Full fine-tuning on Whisper \textit{large} (large-Full-FT) is not conducted due to limited GPU memory.}
  \label{fig:fullft_vs_peft}
  \vspace{-5mm}
\end{figure}

\section{Results and Anaysis}
\label{sec:result}

\subsection{Why we need PEFT on Whisper?}
To prove the necessity of PEFT for Whisper, the performance comparison between Full-FT and PEFT is carried out. AdaLoRA ($r=8$) is used as a representative PEFT method in this experiment. 

As shown in Fig. \ref{fig:fullft_vs_peft}, five different sizes of Whisper are evaluated for Chinese child speech recognition. The x-axis represents the varying amount of adaptation data from zh-C1-1K to zh-C1-30K. The CER in the y-axis indicates the recognition performance on evaluation sets of zh-C1 (left) and zh-C2 (right). 
Full-FT is plotted by solid curves and PEFT use the dashed ones. 

We can notice that Whisper with PEFT clearly outperforms that of using Full-FT on the zh-C1 in-domain evaluation and keeps a similar error reduction trend on zh-C2. 
Particularly for Full-FT, it can be observed only the \textit{tiny} and \textit{base} Whispers show the improvements after adaptation, while the \textit{small} and \textit{medium} all made a large performance degradation, especially for the case of 1K utterances. However, PEFT gives consistent improvements across different model sizes and different amounts of adaptation data. In addition, PEFT on \textit{medium} (a half size of \textit{large}) surpasses a lot over those with smaller model sizes, approaching the level of \textit{large}. If not specified otherwise, Whisper \textit{medium} is utilized in the following experiments.

\begin{table}[t]
\centering
\caption{The CER(\%) performance comparison between AdaLoRA and LoRA on the evaluation sets of zh-C1 and zh-C2, where the adaptation data is zh-C1-1K. $r$ denotes the rank in LoRA or target rank in AdaLoRA. Initial rank of AdaLoRA is set to 48 for $r=32$, otherwise being 12.   The $orth$ and $alloc$ represent the orthogonal regularization and rank allocation in AdaLoRA, respectively. }
\resizebox{0.95\linewidth}{!}{
\begin{tabular}{c|ccc|ccc}
\toprule
\multirow{2}{*}{PEFT} & \multicolumn{3}{c|}{zh-C1} & \multicolumn{3}{c}{zh-C2} \\ \cline{2-7} 
                      & $r$=1      & $r$=8    & $r$=32   & $r$=1     & $r$=8     & $r$=32   \\ \midrule
Zero-shot             & \multicolumn{3}{c|}{11.29} & \multicolumn{3}{c}{19.25} \\ \midrule
LoRA                  & 10.63    & 9.32   & 9.22   & 17.87   & 14.73   & 14.45  \\
AdaLoRA               & 8.93     & \textbf{8.46}   & 8.79   & 14.26   & 13.73   & 13.48  \\
w/o orth              & 8.81     & 8.62   & 8.78   & 14.46   & 12.56   & 13.35  \\
w/o alloc             & 8.53     & 8.53   & 8.78   & 13.80    & 13.80    & 13.38  \\
w/o  both     & 8.64     & 8.64   & 8.76   & 12.52   & 12.52   & 13.41  \\ \midrule
$\alpha$-LoRA                & 8.86     & \textbf{8.77}   & 8.47   & 13.75   & 13.68   & 12.70  \\ \bottomrule
\end{tabular}}
\label{tab:adlora_vs_lora}
\end{table}

\subsection{What makes AdaLoRA superior to LoRA?}
\label{subsec:adalora_vs_lora}
AdaLoRA always gives significantly better PEFT performance than LoRA in our experiments, which is in line with previous findings \cite{DBLP:conf/iclr/ZhangCBH0CZ23}.  As mentioned in section \ref{subsec:adalora}, AdaLoRA mainly comes with two modifications to the vanilla one: (1) orthogonal regularization and learnable rank coefficients via SVD; (2) importance-aware rank allocation. To investigate the main reason that makes AdaLoRA superior to LoRA, the following ablation experiments are performed:(1) remove orthogonal regularization ($orth$) from AdaLoRA; (2) remove importance-aware rank allocation ($alloc$) from AdaLoRA; (3) remove $orth$ and $alloc$ from AdaLoRA, only learnable rank coefficients left; (4) add rank coefficients to LoRA, since rank coefficients used for rank allocation can not be removed from AdaLoRA. As illustrated in Tab. \ref{tab:adlora_vs_lora}, Whisper \textit{medium} was used to adapt on zh-C1-1K by different ablation variants of AdaLoRA. 


It can be seen that three different rank values $\{1,8,32\}$ were tested and AdaLoRA all obviously surpasses LoRA. By removing $orth$, $alloc$ and both of them from AdaLoRA, clear performance degradation does not happen, even bringing some performance gains. This shows that $orth$ and $alloc$ are not the key factors for the performance improvement of AdaLoRA compared to LoRA. 

 
To verify the benefits of learnable rank coefficients, a trainable scalar $\alpha$ was introduced to LoRA, named $\alpha$-LoRA for experiments where $\Delta\mathbf{W}_i = \alpha \mathbf{B}_i \mathbf{A}_i$. For vanilla LoRA, $\mathbf{B}_i$ has to be initialized as zero to make sure $\Delta\mathbf{W}_i=\mathbf{0}$ at the start. With $\alpha=0$, $\{\mathbf{B}_i, \mathbf{A}_i\}$ can be initialized by a normal distribution, similar to the initialization way of AdaLoRA. It can be seen from Tab. \ref{tab:adlora_vs_lora}, $\alpha$-LoRA largely improves the vanilla LoRA across all ranks, illustrating the learnable rank coefficients are the most important design. 


\subsection{Comparing S2-LoRA with other PEFTs}
The proposed S2-LoRA is inspired by the design of rank coefficients (See section \ref{subsec:adalora_vs_lora}) and the sparse rank distribution of AdaLoRA. 
Tab. \ref{tab:s2lora} evaluates S2-LoRA with several other \textit{parameter composition} types of PEFT methods.

AdaLoRA with 0.46\% trainable parameters achieves the best in-domain performance while GLoRA with 0.81\% ones exhibits the best generalization ability. IA3 with 0.03\% trainable parameters performs the worst among PEFT approaches. Compared to AdaLoRA, the proposed S2-LoRA is competitive on in-domain data and performs better under out-of-domain conditions, while using 20x fewer trainable parameters (0.02\%). Notably, S2-LoRA has trainable parameters even fewer than IA3 and BitFit. 

For cross-lingual evaluation, all PEFT methods achieved positive performance gains in Mandarin (zh-C1, zh-C2, and zh-A1), while suffering from increased CERs (around 2-3\%) in Cantonese (zh-HK). This may be due to the conflict in the token modeling of Whisper where the same language ID \textit{zh} is used for both Mandarin and Cantonese. Performance improvement in Mandarin would cause a performance reduction in Cantonese. With different language IDs (\textit{en} vs. \textit{zh}), all PEFT methods are found to be able to bring somewhat improvements in English while adapting to Mandarin.

\begin{figure}[t!]
  \centering
  \includegraphics[width=1.0\linewidth]{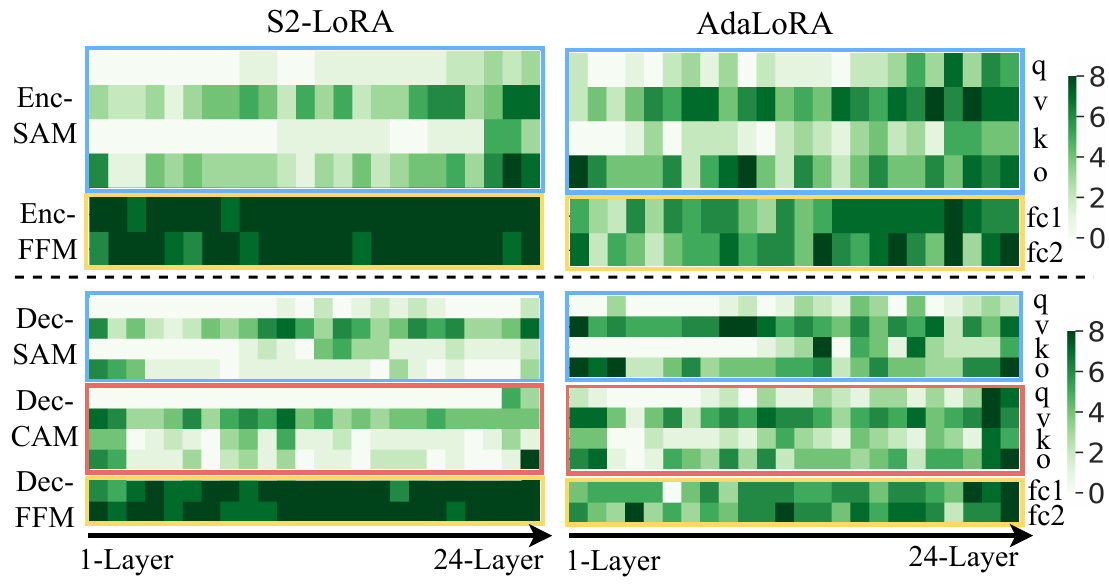}
  \caption{The rank distribution of S2-LoRA ($r=8$) and AdaLoRA. For a fair comparison, the initial rank of AdaLoRA is set to 8, and the target rank is 4. In S2-LoRA, the rank coefficients less than $1e-4$ are masked out. The row order in SAM/CAM follows $\{\mathbf{W}_q, \mathbf{W}_v, \mathbf{W}_k, \mathbf{W}_o\}$, while in FFM follows $\{\mathbf{W}_{fc1}, \mathbf{W}_{fc2}\}$.}
  \label{fig:s2-lora-rank}
  \vspace{-2mm}
\end{figure}

Furthermore, the rank distribution of S2-LoRA is found to have similar patterns with AdaLoRA as shown in Fig. \ref{fig:s2-lora-rank}. Each element of the distribution matrix denotes the allocated rank of the incremental update matrix $\Delta\mathbf{W}_i$. The elements with brighter colors represent lower ranks allocated, thus being less important. 
In the SAM/CAM of both encoder and decoder, the rank allocated to $\{\mathbf{W}_q\, \mathbf{W}_k \}$ are mostly zero, suggesting that they are much less important than $\{\mathbf{W}_v, \mathbf{W}_o\}$.  This phenomenon exists in AdaLoRA and S2-LoRA, indicating that S2-LoRA learns to identify important blocks and allocate ranks to them for improving adaptation performance.

\begin{table}[t]
\centering
\caption{The CER(\%) performance comparison between S2-LoRA and other PEFT methods where rank=8. Whisper \textit{medium} with zh-C1-1K as the adaptation data was tested on in-domain data (zh-C1) and out-of-domain data (others). WER(\%) is used for English (en).}
\resizebox{1.0\linewidth}{!}{
\begin{tabular}{cc|c|cccc}
\toprule
\multirow{2}{*}{Methods} & \multirow{2}{*}{\begin{tabular}[c]{@{}c@{}}\#Trainable\\  Params\end{tabular}} & \multicolumn{1}{l|}{in-domain} & \multicolumn{4}{c}{out-of-domain} \\ \cline{3-7} 
                         &                                                                                & zh-C1                          & zh-C2  & zh-A1  & zh-HK  & en     \\ \midrule
Zero-shot                & -                                                                              & 11.29                          & 19.25  & 9.79   & 31.14  & 7.93   \\ \midrule
Full-FT                  & 100\%                                                                          & 19.86                          & 31.97  & 28.90   & 57.70   & 11.47  \\
GLoRA                    & 0.81\%                                                                         & 8.85                           & \textbf{13.04}  & 6.71   & \textbf{31.79}  & 6.69   \\
AdaLoRA                  & \textbf{0.46\%}                                                                         & \textbf{8.46}                           & 13.73  & 7.65   & 34.79  & 7.57   \\
LoRA                     & 0.31\%                                                                         & 9.32                           & 14.73  & 7.52   & 33.30   & 7.35   \\
BitFit                   & 0.08\%                                                                         & 8.72                           & 13.38  & 7.85   & 34.77  & 7.50   \\
IA3                      & 0.03\%                                                                         & 10.24                          & 15.73  & 9.55   & 34.26  & \textbf{6.62}   \\ \midrule
$\alpha$-LoRA                   & 0.31\%                                                                         & 8.77                           & 13.68  & \textbf{6.37}   & 33.01  & 6.73   \\
S2-LoRA                  & \textbf{0.02\%}                                                                         & \textbf{8.67}                           & 13.19  & 6.81   & 33.71  & 6.72   \\ \bottomrule
\end{tabular}}
\label{tab:s2lora}
\end{table}

\section{Conclusions}
\label{sec:conclusion}
This paper presents a novel S2-LoRA approach to adapting Whisper. S2-LoRA improves AdaLoRA with much fewer trainable parameters than most existing PEFT methods such as GLoRA, BitFit, etc. It benefits from the design of the sparse learnable rank coefficients and shared rank-decomposed matrices. 
Experiments carried out on low-resource Chinese child speech demonstrate the effectiveness of our proposed approach, showing that S2-LoRA achieves adaptation performance comparable to AdaLoRA and noticeably better cross-domain generalizability while retaining only $0.02\%$ trainable parameters. Though using child speech as a study case, the proposed S2-LoRA is general and can benefit adaptation in other low-resource speech recognition scenarios. 


\bibliographystyle{IEEEbib}
\bibliography{refs}

\end{document}